# Does Search Engine Optimization come along with high-quality content?

SEO and content quality

## A comparison between optimized and non-optimized health-related web pages

Sebastian Schultheiß

Hamburg University of Applied Sciences, sebastian.schultheiss@haw-hamburg.de

Helena Häußler

Hamburg University of Applied Sciences, helena.haeussler@haw-hamburg.de

Dirk Lewandowski

Hamburg University of Applied Sciences, dirk.lewandowski@haw-hamburg.de

Searching for medical information is both a common and important activity since it influences decisions people make about their healthcare. Using search engine optimization (SEO), content producers seek to increase the visibility of their content. SEO is more likely to be practiced by commercially motivated content producers such as pharmaceutical companies than by non-commercial providers such as governmental bodies. In this study, we ask whether content quality correlates with the presence or absence of SEO measures on a web page. We conducted a user study in which $N = 61$ participants comprising laypeople as well as experts in health information assessment evaluated health-related web pages classified as either optimized or non-optimized. The subjects rated the expertise of non-optimized web pages as higher than the expertise of optimized pages, justifying their appraisal by the more competent and reputable appearance of non-optimized pages. In addition, comments about the website operators of the non-optimized pages were exclusively positive, while optimized pages tended to receive positive as well as negative assessments. We found no differences between the ratings of laypeople and experts. Since non-optimized, but high-quality content may be outranked by optimized content of lower quality, trusted sources should be prioritized in rankings.

**CCS CONCEPTS** •Information systems~World Wide Web~Web searching and information discovery~Web search engines

**ACM Reference Format:**

# 1 Introduction

Search engines are an integral part of our everyday lives and among the most frequently used services on the Internet [17,36]. The ubiquity of search engine use also applies to health-related topics. In 2020, more than one in two EU citizens (55%) searched online for health-related information. This is a sharp increase compared to 2010, when this figure was only 34% [10]. In a recent survey by search engine optimization analytics company

moz.com, 46% of respondents indicated that they often or very frequently use Google to make important medical decisions [37]. Thus, if a result is noticed and selected on a search engine result page (SERP) due to its position [15,34], visibility [6,39], or design [31], this impacts decision making. When users make decisions based on a limited set of results chosen because of their position or attractiveness, it can also affect society beyond public health, for instance when users research legal, financial, [37] or political matters online. For example, Epstein and Robertson [8] showed that when search engines favor results with positive statements about a political candidate on their SERPs, they can influence user voting preferences. Applied to the context of online health information-seeking behavior (OHISB) [25,54], this could, for example, result in preferences for certain types of treatments or medications simply because the corresponding results are ranked higher on the list.

Search engines are not always capable of ranking the best results first (here, best is defined as results that provide a user with the most reliable, comprehensive, and up-to date information). For instance, it has been shown that the top search results are biased towards commercial interests when comparing insurance offers [28], while other studies show the opposite regarding health information [e.g., 31]. In addition, search engine operators often preselect trusted sources, for example when searching for COVID-19. It is however unclear how, why, and by whom these decisions were made [47]. It is evident that a combination of user selection behavior and sub-standard search results can have a direct impact on the everyday lives of search engine users [35]. This impact ranges from wasted time as a result of irrelevant results to users choosing inappropriate or even harmful medical treatments. Several factors influence where a particular result is displayed on the SERP. These include search engine providers, users, paid search marketing (PSM), which refers to keyword-related advertisements, content producers, and search engine optimization (SEO) as shown in Figure 1 [38].

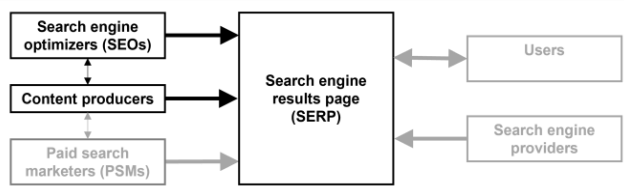

**Figure 1: Stakeholder groups that influence results (based on** [38]**)**

For any website that seeks to achieve visibility, SEO has become a necessary standard method of online marketing [16,43]. SEO is "the practice of optimizing web pages in a way that improves their ranking in the organic search results" [30]. SEO is therefore a kind of reverse engineering of search engine ranking algorithms with the aim of exploiting knowledge about ranking methods for the benefit of one's own information objects [30]. For 2021, total sales of 50 billion dollars are forecast for the global search engine optimization services market, although this estimate has been affected by COVID-19. The market is expected to recover from the impact of the pandemic and grow to 103 billion in 2025 [48]. Despite the size and importance of the SEO industry, most users are not familiar with search engine optimization. As survey research has shown, most internet users are not aware of the existence of search engine optimization, the term "SEO," or SEO measures. Furthermore, most users do not associate SEO with organic results. The small portion of users to whom this does not apply and who have greater knowledge of SEO are typically younger, highly educated, and work in a SEO-related profession [27].

In this article, we focus on SEO and content producers, which is why both areas are highlighted in **Fehler! Verweisquelle konnte nicht gefunden werden.**. Due to their business models, the SEO industry and content producers are directly dependent on one another [43]. SEO measures are carried out either by the content producers themselves or by hired SEO agencies [38] to ensure content reaches users, a tactic Mager [32] also observed among health information providers. The motivation for conducting SEO activities can be commercial and short-term, e.g., to boost sales, but also non-commercial and more long-term or political in nature [38]. How users perceive the intention of health-related websites affects how they think about those websites. Whether a website is perceived as impartial, for instance through the absence of advertising, directly affects user trust in health-related websites [18,40]. Therefore, a connection between SEO, commercial interests, and the quality of search results can be deduced, since SEO can be seen as an expression of the content producers' predominantly commercial interests, and SEO measures can improve the ranking even of lower-quality content. This leads us to the question of whether the presence or absence of SEO measures allows conclusions to be drawn about content quality of those web pages. Since SEO has a measurable influence on result rankings [29] and thus on what users see, we must also ask how users are affected by the probable higher rankings of optimized content.

We sought to determine whether quality differences are perceived between optimized and non-optimized web pages. To accomplish this, we conducted a user study consisting of web page evaluations, thinking aloud protocols, and surveys. We invited laypeople and experts in health information assessment to evaluate the quality of optimized and non-optimized health-related web pages. Quality is determined as trustworthiness, expertise, objectivity, transparency, familiarity, understandability, and relevance related to the web page. For this purpose, we made use of an SEO classification tool. Based on $N = 21$ indicators, a rule-based classifier determines the probability of SEO on a web page by using its HTML code and further indicators [29]. In addition to asking whether there are differences between optimized and non-optimized web pages, we also address the question of how these differences are perceived by means of thinking aloud. At the end of the study, the subjects were given short surveys to ascertain their professional backgrounds and elicit their opinions and knowledge about SEO. Our goal is to gain a better understanding of the potential impact SEO may have on the information seeking process and decision-making of search engine users when it comes to health-related problems.

The rest of the paper is structured as follows. First, we provide an overview of studies on search result perception and selection, information literacy of search engine users, and user evaluation of online health information. Next, we state the research questions and describe the methods used. After describing how we analyzed the data, we present the results in six parts. Finally, we discuss the results, provide a summary, and present suggestions for future research.

## 2  Literature review

### Perception and selection of search results

SEO is of considerable relevance when we consider how users interact with search results. In general, they prefer to view and click top-ranked search results regardless of relevance [1,20,22]. The variety of websites represented in the top results is however limited. A large-scale study showed that just 10,000 unique websites accounted for approximately 80 percent of the targets of clicked results for the 2.6 billion queries evaluated in the study [14]. Furthermore, users tend to select search results that confirm their prior beliefs [49]. In terms of health information, they gravitate toward positive results [35,49]. Most queries are formulated positively, which leads to content biased towards results that *suggest* that an intervention helps [51]. Attention is also directed towards results with alarming captions, for instance describing serious symptoms or diseases. Regardless of their position in the ranking, such results are clicked more frequently and examined in greater detail [52]. This attention bias may reinforce concern or anxiety, particularly for users who were concerned *prior* to searching [42]. Another factor is device screen size, which affects result perception and selection. Users focus on the visible area of the SERP that is accessible without scrolling [19] and mainly select results from that area [23]. This search behavior demonstrates that multiple factors are influencing users. SEO measures may leverage and reinforce certain selection behavior and biases.

### Information literacy of search engine users

Users are, for the most part, confident in their interactions with search engines. According to an EU-wide survey, subjects say they usually find what they are looking for and believe advertisements on the SERP are clearly recognizable [9]. However, the self-assessments of search skills seem exaggerated when tested in task-based studies. Lewandowski et al. [26] found that users possess little knowledge of search engines and lack the ability to distinguish between advertisements and organic results—despite optimistic self-assessments. A recent study [27] confirms this discrepancy and also finds limited knowledge of SEO and the fact that it influences search results. This inadequate knowledge of ranking mechanisms and business models stands in contrast to the awareness and application of SEO measures by the operators of health websites [32]. Highly professional content producers who know how to influence results pages through SEO activities thus coexist with users who are unaware of these possibilities. Additionally, due to a lack of domain knowledge, laypeople face challenges when attempting to formulate search queries that accurately represent their information needs [24]. Consequently, users may select irrelevant websites to confirm their incorrect initial hypotheses [24]. Laypeople also tend to select websites with possible commercial motivations (.com), whereas experts rely on educational (.edu) or governmental (.gov) sites [5,50]. Information literacy has the potential to compensate to some extent for the laypersons' missing domain knowledge [24].

In summary, users are exposed to a diverse set of influences and interests largely in the absence of theoretical and practical knowledge of search engines that would enable them to critically evaluate the retrieved information.

## User evaluation of online health information

Online resources increasingly shape health information-seeking behavior (HISB) [54] with search engines constituting the starting point of health information research for a majority of citizens [2,10,11]. According to a survey representative of the German population, evaluating the quality of health information is challenging for laypeople. When asked about issues concerning coping with illness, about half of respondents (49%) said the trustworthiness of health information was difficult to assess, followed by those who found it difficult to weigh the advantages and disadvantages of certain forms of treatment (45%) and judging when to seek a second opinion (42%) [41]. Although several initiatives have created certificates in an attempt to foster the publication of high-quality health websites, checklists, and guidelines [3,7,21], the industry still lacks a widely recognized standard. Sun et al. extensively reviewed studies that examine the criteria users apply when assessing the information quality of online health information [46]. They found trustworthiness to be the most frequently mentioned criterion, followed by expertise and objectivity. The studies also revealed that features of website design and information content can influence trust judgments positively or negatively [40]. For instance, a clear layout and a content producer with discernible competence contribute considerably to trust building, whereas advertisements have the opposite effect [40,46]. Visual cues are important for first impressions but are overshadowed by content features over the course of frequent use [45].

In short, information quality plays an essential role in the evaluation of online health information. Laypeople may be sensitive to differences in quality and prevailing interests at the level of the website. Evaluation strategies focus on content, source, and design features. At the level of the SERP, users seem to be more susceptible to external influences.

## 3  Research questions

In this section, we provide the research questions. RQ1 focuses on differences between optimized and non-optimized web pages. To consider whether professional handling of health-related websites might influence ratings of health-related websites, we distinguish between experts and laypeople regarding the evaluation of health information (RQ2). We also collect justifications for the ratings given (RQ3; see also thinking aloud protocols in methods section). As mentioned in the introduction, a small portion of users has greater knowledge and more positive opinions of SEO. This is addressed by RQ4, which seeks to determine the extent to which web page evaluations differ between groups with divergent attitudes and knowledge levels with regard to SEO.

**RQ1**: What differences can be found in the quality assessments of health-related web pages when the web pages are differentiated by presence/absence of SEO?

**RQ2**: What are the differences between the evaluations of laypeople and those of experts?

**RQ3**: How do laypeople justify their evaluations of optimized and non-optimized web pages?

**RQ4**: How do SEO knowledge and the attitudes of laypeople influence their evaluation of the optimized and non-optimized web pages?

## 4 Methods

We conducted a user study in April and May of 2021 consisting of web page evaluations, think aloud protocols, and surveys. Due to the Corona pandemic, the study was conducted online, using the survey software *EFS Survey by Questback* (accessed via the portal *Unipark*)[1] and *Zoom Meetings*[2].

### 4.1 Procedure

Two groups were distinguished, each of which completed separate project parts: study part laypeople and study part experts. As shown in Figure 2, the laypeople completed the evaluation of health-related web pages combined with thinking aloud protocols (Stage 1), followed by a survey (Stage 2).

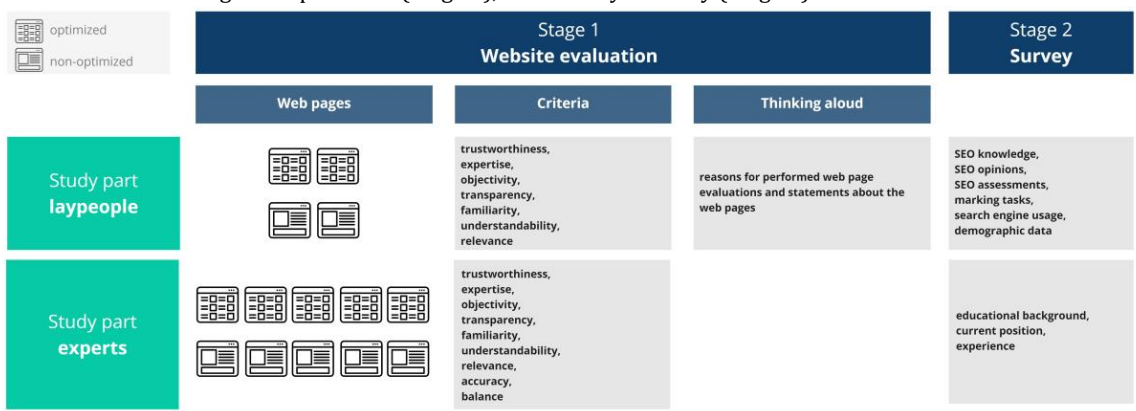

**Figure 2: Study parts**

In Stage 1, each layperson evaluated two optimized and two non-optimized web pages in randomized order for health-related queries described in Section 4.2. Since each layperson received different web pages and each web page was evaluated once, a total of *N* = 200 pages were evaluated by the laypeople. We have placed the following explanation above each web page, with the name of the disease (e.g., asthma) matching the web page content in each case:

> Imagine you are looking for information on the disease asthma and are shown the following web page. Please look at the web page and consider how you would rate its quality.

Below the explanation, the web page was shown as a scrollable screenshot. Underneath the screenshot, we placed sliders to evaluate criteria of information quality. The laypeople evaluated the web pages for the criteria of trustworthiness, expertise, objectivity, transparency, familiarity, understandability, and relevance. We chose these criteria for two reasons. First, these criteria are among the most frequently chosen by users when assessing the quality of online health information [46]. Second, we consider these criteria to be assessable both in terms of their meanings and in terms of their quantity for the laypeople. For the sliders we used 101-point scales with labeled endpoints ("not at all," "extremely"). The definitions for the criteria [46] were formulated as questions to assist subjects in case they were doubtful about the meaning of a criterion. They could be accessed through a question mark button. While evaluating the web pages, laypeople commented on their actions by thinking aloud. The statements were recorded within *Zoom Meetings*. This allowed us to capture the immediate impressions subjects had regarding the web pages, which they may not have fully remembered in the survey at the end of the study. After completing the web page evaluation and thus Stage 1, a survey followed in Stage 2. The questions were derived from [27]. The questionnaire comprises the following sections: (1) knowledge of SEO, (2) ability to assign results influenceable by SEO/PSM (marking tasks), (3) assessments and opinions regarding SEO, (4) search engine usage, and (5) demographic data. See Appendix 1 for the complete questionnaire. In Section 2 (marking tasks), the laypeople viewed screenshots of Google results pages with one SERP having a simple structure consisting of organic results and text ads only and one SERP being more complex and comprising organic results, text ads, shopping ads, and news results. On the screenshots, the laypeople had to mark either all results that are paid (i.e., ads) or all results that are influenceable by SEO (i.e., organic results). Using the correct and incorrect markings, success rates were calculated for each task. The higher the success rate, the more precisely a subject identified the appropriate result type. For more information on how the marking tasks were designed and the success rates calculated, see [44].

---

[1] https://www.unipark.com/en/survey-software/
[2] https://zoom.us/meetings

The portion of the study that involved experts was questionnaire-based only and thus conducted without thinking aloud protocols. Since we acquired only a limited number of experts, not all web pages that the laypeople evaluated ($N = 200$) could also be evaluated by the experts. The experts assessed a subset ($N = 110$) of the web pages using the same criteria as the laypeople. They additionally evaluated the web pages in terms of accuracy and balanced presentation of content. Each expert considered five optimized and five non-optimized web pages (stage 1, see Figure 2). Afterwards, the experts received a short survey covering their professional backgrounds (stage 2). After participating, all subjects received a detailed information sheet about the objectives of the study and a 10 EUR Amazon voucher.

## 4.2 Sampling and classification of web pages

In the following, we will refer to both automatic and manual classifications. The former refers to the classifications our SEO classification tool performs, the latter to the manual classification according to web page type. The sampling of the web pages was done in three steps: (1) We automatically identified optimized and non-optimized web pages. (2) We then manually assigned the web pages to different web page types, e.g., journalistic or commercial content. (3) We used these web pages as the samples for our study.

In the first step, we performed an analysis of the results to $N = 15$ health-related queries using our SEO classification tool [29] in February 2021. These queries relate to broadly known diseases, e.g., acne, epilepsy, and cataract, and were taken from gesund.bund.de, a service of the German Federal Ministry of Health.

Among $N = 1,059$ results[3] generated by the queries, the SEO classification tool automatically classified $N = 337$ web pages as optimized and $N = 200$ pages as non-optimized[4].

For this purpose, the tool first submitted the queries to Google, collected the result URLs and result positions from the SERPs, collected the result documents, and analyzed the HTML code. Based on $N = 21$ indicators, a rule-based classifier determined the likelihood of SEO on a web page. A web page was classified as optimized if the intention to perform SEO was clearly recognizable, e.g., a SEO plugin was discovered on the analyzed web page, or if a web page is journalistic in nature[5], as SEO is an essential tool to reach readers. This applies to content producers in general and journalistic content in particular according to interviews with journalists [13,43]. A web page was classified as non-optimized if it did not meet basic SEO criteria, such as the use of a title or meta description tag [29]. Second, the optimized and non-optimized web pages ($N = 537$ in total) were manually assigned to different web page categories. The classification was conducted by a student assistant and was derived inductively from the web pages. The web page types are as follows:

---

[3] To ensure efficient data collection, we let the tool evaluate approximately the first seven SERPs for each query.
[4] Since we decided to only consider pages clearly classified as optimized or non-optimized by our tool, we excluded results from Wikipedia ($N = 15$) and pages classified as "probably optimized" ($N = 507$) from the analysis.
[5] For this purpose, the documents were matched against a list of manually classified journalistic websites as described in [29].

- Journalistic content
- Doctor's office, clinic
- Insurance company
- Information, commercial (e.g., pharmaceutical companies)
- Information, non-commercial (e.g., university hospitals)
- Information offered by a public authority (e.g., ministries)
- Other (pharmacy, shop, video)

In the third step, two samples of optimized and non-optimized web pages were drawn to be evaluated by the subjects, each consisting of $N$ = 100 web pages. For all web pages, we created screenshots using the desktop version of the Chrome browser at a zoom level of 150% to optimize the readability of the screenshots for the survey tool. We used a browser extension for screen capturing[6] to create full-size web page screenshots. Research data is available via (https://osf.io/uk4eq).

## 4.3 Sampling of laypeople and experts

As described in section 3, we recruited laypeople and experts to participate in our study. We invited $N$ = 50 laypeople, distinguishing between two subject groups, both consisting of $N$ = 25 subjects: A) greater knowledge of SEO and B) lesser knowledge of SEO. To form these groups, we used characteristics from a representative online survey on SEO [27] that correlate with higher (young age, high level of education, and topic-related studies, e.g., information science) or lower knowledge of SEO. Based on these characteristics, we gradually invited laypeople, analyzed their questionnaire results on the SEO-related questions and marking tasks, and then assigned them either to group A (correct answers to most of the SEO-related questions *and* success in SEO marking tasks[7]) or group B (incorrect answers to the SEO questions *or* failure in SEO marking tasks). The laypeople were acquired via university e-mail distribution lists, eBay classified ads, a neighborhood community portal, newspaper ads, Facebook Marketplace, and a forum for elderly people.

In addition to the laypeople, we sampled $N$ = 11 medical librarians, hereinafter referred to as "experts," since these individuals possess both expertise in evaluation of information and in handling medical/health information. Medical libraries were identified through desk research and requested directly and via a mailing list of the working group for academic medical libraries[8].

## 5 Data analysis

## 5.1 Transcription and Coding

First, the think-aloud comments from laypeople assessing the web pages were transcribed. Transcriptions were classified according to whether they represented a positive or negative assessment. Transcribed comments were additionally linked to the criteria from [46], namely, trustworthiness, expertise, objectivity, transparency, familiarity, understandability, and relevance. Laypeople either directly justified their ratings of the criteria (e.g., by saying "the site seems *transparent* because ..."), or the connection was drawn by us in cases in which the layperson did not explicitly link their utterances to the criteria. For this purpose, we used the indicators introduced by [46] for the individual criteria (e.g., "disclosure of the site owner" as a positive indicator for the criterion transparency). In a second step, the transcribed utterances were coded. The codes were created inductively based on the text and in total encompassed $N$ = 52 codes, most of them affecting elements either of the web pages, the texts, or the website owner and authors. Coding was carried out by one researcher and checked by a second. Inconsistencies were discussed and agreed upon. For the coding of the open answers of the laypeople questionnaire, see Appendix 1.

---

[6] https://chrome.google.com/webstore/detail/gofullpage-full-page-scre/fdpohaocaechififmbbbbbknoalclacl
[7] Success in the SEO marking tasks (i.e., the marking of organic results; see section 4.1) we defined as a score that is higher than the median of all the laypeople on the SEO-related marking tasks.
[8] A working group of the German Medical Library Association (https://www.agmb.de/de_DE/arbeitskreise#arbeitskreis-medizinbibliotheken-hochschulen)

## 5.2 Statistical methods

To test whether web page optimization or group affected the evaluation of web page quality (section 6.4), we conducted a repeated measures analysis of variance (ANOVA). We defined optimization of a web page (optimized/non-optimized) as the within-subjects variable and group (laypeople/expert) as the between-subjects factor. We conducted separate ANOVAs for each criterion individually. In section 6.3 we compare the positions of the optimized with those of non-optimized pages by conducting a two-sample t-test. In section 6.5 we first compare two groups of SEO knowledge (less knowledge vs. greater knowledge) by conducting Welch's t-test for unequal variances. Second, since the questions regarding the assessments of the degree of SEO influence and SEO opinions consist of more than two groups (i.e., response options), single-factor analyses of variance (ANOVAs) were performed for each question.

# 6 Results

We describe the results in six parts. First, the characteristics of the subjects are presented. Second, we address the laypeople questionnaire results with respect to SEO knowledge and attitudes. Third, we present the results of the manual classification of optimized and non-optimized web pages and fourth, how these web pages were evaluated by the subjects. Fifth, we show what influence the SEO knowledge and attitudes of laypeople have on their web page evaluations. Finally, the thinking aloud results regarding the rationales for the evaluations are presented.

## 6.1 Characteristics of laypeople and experts

We acquired $N$ = 50 laypeople with a mean age of 36.34 years (SD = 15.76; range between 18 and 68 years), 58% of them were female, 42% male. As their highest educational level, the laypeople indicated A-levels (university entrance qualification; 46%) followed by university degree (42%), General Certificate of Secondary Education (10%), and other (2%). Most laypeople were currently in training or apprenticeship (54%), followed by being an employee or public official (32%), not or no longer employed (10%), or self-employed, freelancer, or entrepreneur (4%). Among the laypeople, Google is the most-used search engine with a 94% use rate. 80% use it most frequently while also using other search engines. All laypeople use search engines to research health information, about one third (34%) do so once a month, followed by several times per month (28%) and less than once a month (24%).

All $N$ = 11 experts have an educational background in Library and Information Science, while one person additionally has training and professional experience in nursing and the humanities. The experts hold a wide range of positions, i.e., leading positions in the library ($N$ = 3), librarians ($N$ = 3), subject librarians ($N$ = 2), information specialist ($N$ = 1), library assistants ($N$ = 1), or student assistants ($N$ = 1). Regarding the years of professional experience, the experts can be divided into newcomers with experience in the field of medical libraries of up to 5 years ($N$ = 6) and senior experts who have been affiliated with the sector for more than 5 years ($N$ = 5). In their professional lives, the experts are frequently confronted with websites containing health information. Half of them deal with health websites on a daily or weekly basis ($N$ = 5), the other half seldom work with health websites, only a few times per month or less ($N$ = 6).

## 6.2 SEO knowledge and attitudes of laypeople

The following results refer to the questions on SEO (Questions 1.1-1.3), the marking tasks (Questions 2.1-2.4), and the questions regarding assessments and opinions of SEO (Questions 3.1-3.4) as shown in Appendix 1. Of all laypeople, 80% assumed that improved rankings can be achieved without paying money to Google (Question 1.1).

The other laypeople answered either incorrectly ("No"; 6%) or "I don't know" (14%). Among the laypeople who answered "Yes" to question 1.1, 68% said they knew the term that is used to describe these measures (Question 1.2). However, only 46% correctly named "SEO" or "search engine optimization" (in English or German). Among all laypeople, 62% said they knew SEO techniques (question 1.3), while 58% correctly named techniques such as keyword optimization or backlinks.

We will now consider the results of the marking tasks. As Table 1 shows for the simple structure SERP, success in identifying PSM and SEO results[9] was roughly equivalent (.53 vs. .58). A different picture emerges for the complex structure SERP. Here, ads (PSM) were identified with greater accuracy (.58) than organic results (SEO) (.29).

**Table 1: Mean success rates (marking tasks)**

|  | Success rate: PSM, mean (SD) | Success rate: SEO, mean (SD) |
|---|---|---|
| Simple SERP | .53 (.78) | .58 (.43) |
| Complex SERP | .58 (.43) | .29 (.54) |

Most laypeople assumed that SEO has a strong (44%) or very strong (40%) influence on ranking. Only a few respondents assumed medium (10%) or little (6%) influence (Question 3.1). Laypeople more often indicated (very) large positive (60% overall) than (very) large negative (40% overall) SEO influence (Question 3.2).

## 6.3 Manual web page classification

In this section, we present the results of the manual web page classification. Figure 3 shows the classifications of the optimized ($N = 337$) and non-optimized ($N = 200$) web page types as described in Section 4.2.

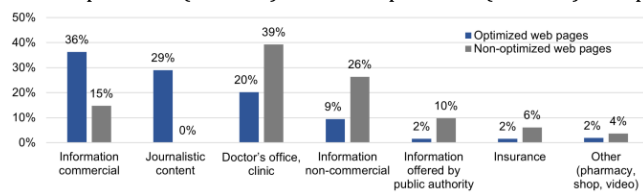

**Figure 3: Types of optimized and non-optimized web pages**

When looking at optimized web pages, commercial information (provided by pharmaceutical companies, for instance) accounts for the largest share (36%). This is followed by journalistic content (29%), which is hardly surprising since the SEO classification tool automatically classifies news sources as optimized (as described in section 4.2). In contrast, the commercial information category includes far fewer non-optimized web pages (15%) than optimized web pages. Non-optimized web pages are most prevalent among doctor's offices and clinic web pages (39%), followed by non-commercial information (26%), and information provided by public authorities (10%) such as government departments and agencies.

Since our SEO classification tool also collected the result positions (see 4.2), we compared the positions of the optimized with those of non-optimized pages. The 337 optimized pages ($M = 35.01$, $SD = 20.488$) compared to 200 non-optimized pages ($M = 40.72$, $SD = 19.408$) demonstrated significantly higher ranks on the SERP, $t(535) = 3.184$, $p = .002$.

---

[9] We refer to the organic results as "SEO results." These were to be marked in the marking tasks, insofar as the task asked for search results that are influenceable by SEO measures. See also questionnaire in Appendix 1, section II.

## 6.4 Web page evaluation

We will now look at the laypeople and expert evaluations of samples consisting of optimized and non-optimized web pages (laypeople: $N$ = 200 web pages, experts: $N$ = 110 web pages; see section 4.1). Table 2 presents the evaluations, showing the mean values and standard deviations for laypeople and experts for each criterion.

**Table 2: Evaluation of optimized and non-optimized web pages**

|  | Optimized | | Non-optimized | |
| --- | --- | --- | --- | --- |
|  | Laypeople, N = 100 web pages | Experts, N = 55 web pages | Laypeople, N = 100 web pages | Experts, N = 55 web pages |
| Criterion | mean (SD) | mean (SD) | mean (SD) | mean (SD) |
| Trustworthiness | 70.93 (26.34) | 69.13 (25.23) | 75.87 (23.50) | 78.75 (18.93) |
| Expertise | 69.74 (28.92) | 67.05 (28.35) | 75.81 (24.14) | 76.98 (23.06) |
| Objectivity | 64.66 (26.20) | 61.40 (28.94) | 67.74 (26.33) | 66.78 (27.17) |
| Transparency | 62.70 (30.57) | 60.24 (30.43) | 68.64 (28.24) | 68.33 (29.27) |
| Familiarity | 40.62 (36.58) | 55.05 (36.24) | 39.44 (35.95) | 45.11 (33.36) |
| Understandability | 77.38 (25.24) | 77.98 (20.47) | 76.04 (23.68) | 75.42 (23.04) |
| Relevance | 70.00 (29.18) | 59.76 (32.42) | 70.30 (29.87) | 66.58 (28.16) |
| Accuracy | not evaluated | 69.38 (24.50) | not evaluated | 71.00 (27.31) |
| Balance | not evaluated | 59.55 (29.48) | not evaluated | 59.56 (28.23) |

Conducting separate ANOVAs for all criteria, we only found significant results for the criterion *expertise*. The evaluations for *expertise* differed significantly between optimized and non-optimized web pages [$F(1, 59)$= 4.217, $p$ = 0.044]. However, there was no significant effect of the between-subjects factor "group" [$F(1, 59)$= .026, $p$ = .872]. In summary, whether a web page is optimized or not has a significant influence on the perceived expertise of that web page. Non-optimized web pages are considered more competent than optimized web pages. This is independent of whether the assessment is made by a layperson or an expert. As the manual classification of the web pages indicates (section 6.3), evaluations of non-optimized web pages mainly refer to web pages where no commercial motivation is assumed, such as websites published by public authorities.

## 6.5 Influence of laypeople SEO knowledge and attitudes about web page evaluation

Table 3 shows the web page evaluations by laypeople with lesser and greater SEO knowledge levels (for grouping, see 4.3), differentiated by optimized and non-optimized web pages. For both the non-optimized ($p$ = .843) and optimized ($p$ = .616) web pages, we found no significant differences in evaluation regarding laypeople SEO knowledge.

**Table 3: Web page evaluations of laypeople with different levels of SEO knowledge**

|  | SEO knowledge | Mean | n | SD |
| --- | --- | --- | --- | --- |
| Non-optimized web pages | Lower | 68.08 | 25 | 13.65 |
|  | Higher | 67.30 | 25 | 14.00 |
| Optimized web pages | Lower | 66.16 | 25 | 15.62 |
|  | Higher | 64.14 | 25 | 12.48 |

Looking at the assessed SEO influence strength, we found no significant effect on laypeople evaluations of non-optimized ($F(3,46)$ = .451, $p$ = .718) and optimized ($F(3,46)$ = .296, $p$ = .828) web pages. Regarding positive SEO opinions, there also was no significant effect on evaluations of non-optimized ($F(5,44) = 1.058$, $p = .397$) and optimized ($F(5,44) = .508$, $p = .768$) web pages. The same holds true for the effect of negative SEO opinions on the evaluations of non-optimized ($F(3,46) = .973$, $p = .414$) and optimized ($F(3,46) = .195$, $p = .899$) web pages. In summary, we did not find evidence that SEO knowledge, SEO assessments, or SEO opinions of laypeople influence their evaluations of optimized and non-optimized web pages.

## 6.6 Laypeople justifications of web page evaluations

During the web page evaluations, laypeople were asked to comment on their impressions and give reasons for their assessments. In total, $N = 816$ statements were uttered, whereas $N = 421$ relate to optimized and $N = 395$ to non-optimized web pages. The majority of utterances support one of the seven rating criteria (see procedure described in section 4.1), and 25 % of the comments refer to further criteria listed by [46], such as aesthetics or readability. Apart from distinguishing comments on optimized and non-optimized web pages, a distinction was made between positive and negative statements, i.e., being supportive of or detracting from the evaluation criteria. This results in "pairs of contrast" as many statements concern features which may be interpreted positively or negatively. For example, indicated expertise of the author or disclosed contact information may be received positively when provided or assumed and make a negative impression when absent or assumed to be not provided. Since laypeople evaluated four web pages each, they had the opportunity to compare web pages and focus on similar features.

In the following, the statements associated with the criterion "expertise" are reviewed. Figure 4 shows a comparison of the statements, with the positive mentions (agreeing with the coded statement) on the right and the negative mentions (disagreeing with the statement) on the left. The blue bars indicate statements for optimized web pages while orange bars are for non-optimized web pages. First, it is noticeable that positive comments have been made much more frequently than negative ones.

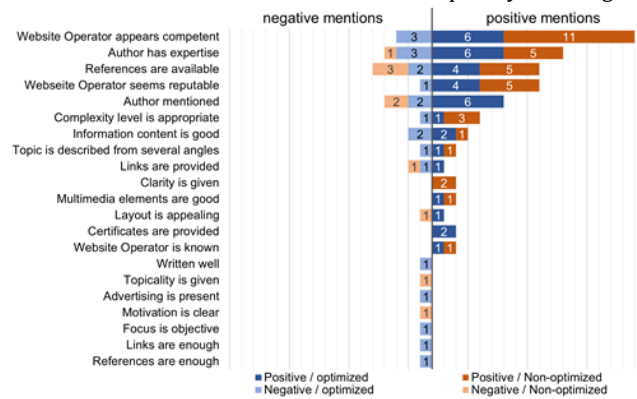

**Figure 4: Statements on the criterion "expertise"**

For both optimized and non-optimized web pages, the most frequent mentions relate to website operators and authors. Users frequently make inferences about a website operator based on the web page's appearance. These include competence and reputation—which may be quite subjective. Looking at the top bar, it is a striking that the website operators of non-optimized web pages received *exclusively* positive comments related to their competence ($N = 11$), whereas optimized web pages, in addition to positive comments ($N = 6$), also received several negative comments ($N = 3$). The most frequent negative mentions for optimized and non-optimized pages combined can be observed for "references." When missing, this was criticized by laypeople five times. Unique positive and negative utterances highlight and contrast specific features among the web pages. The presence of certifications like the HON Code of Conduct ($N = 2$) was mentioned as a positive for optimized web pages only.

Negative aspects attributed to optimized web pages included the presence of advertisements ($N = 1$), an apparent non-objective focus ($N = 1$), and poor quality of the text (negative mention for "well written", $N = 1$). The clear structure of the web page was stressed positively for non-optimized web pages ("clarity is given", $N = 2$). Negative statements related to non-optimized web pages pointed to the lacking topicality of the text ($N = 1$) and to the unclear motivation of the web page ($N = 1$).

We will now briefly present the most frequent statements related to the other criteria. Please note that in several instances, indicators (such as "Website operator looks reputable") were associated with more than one criterion (expertise and trustworthiness, for instance) by the laypeople. The statements on the "trustworthiness" criterion are similar to those on "expertise" in terms of the more positive perception of non-optimized web pages. For optimized web pages, a total of $N = 20$ statements were made that the website operator appears reputable or competent (positive), and $N = 6$ statements expressing that this is not the case (negative). For non-optimized web pages, the ratio is $N = 31$ positive statements to only one negative statement. Regarding the criterion of objectivity, the most frequent complaint about optimized pages is that advertisements were displayed ($N = 16$). The absence of advertisements, on the other hand, was the most frequent *positive* comment about non-optimized web pages ($N = 5$). With respect to transparency, references to medical literature as well as information about the website operator including contact details ($N = 34$ in total) received positive comments, whereas the lack of these elements was criticized as being non-transparent ($N = 18$). Interestingly, for the "familiarity" criterion, rather obvious positive statements such as well-known and reputable-looking website operators ($N = 16$) are contrasted with negative statements again regarding missing literature references ($N = 7$). From the laypeoples' point of view, missing references are therefore problematic for *several* criteria (in addition to familiarity, this also applies to expertise, transparency, and trustworthiness). In terms of understandability, the most frequent comments refer to both structural aspects, i.e., a clear structure of the web page, and content-related aspects, such as the complexity level and comprehensibility of the content. A clear structure was more often attributed to optimized web pages ($N = 12$ positive vs. $N = 2$ negative statements) than non-optimized web pages ($N = 8$ vs. $N = 5$). When it comes to the criterion of relevance, the statements about optimized and non-optimized web pages are quite similar. Good information content ($N = 12$ positive statements) stands in contrast to remarks about too little or too much information ($N = 11$ negative statements).

# 7 Discussion

A key result of our study is that whether a web page is optimized or not has a significant influence on the perceived *expertise* of that page. A higher level of expertise was attributed to web pages classified as non-optimized than to optimized web pages, whereby the latter are ranked higher on the SERPs. As the manual classification showed for the non-optimized web pages, commercial motivations are much less prominent here than with the optimized web pages. No significant differences were found for the evaluation of other criteria such as *trustworthiness* (RQ1). Regarding RQ2, we found no differences between the ratings of laypeople and experts. In terms of the justifications of the evaluations for the criterion of *expertise* expressed through thinking aloud, most utterances relate to features of the source and the expense of content and design. We noticed that laypeople relied both on their subjective impression of competence and reputation and on objective criteria including author expertise and references. Strikingly, website operators of the non-optimized web pages were exclusively considered to be positive (competent, reputable), whereas optimized web pages received both positive and negative comments (RQ3). In addition, we found no connection between the web page evaluations of laypeople and their responses to the SEO-related questions on SEO knowledge, SEO assessments, and SEO opinions (RQ4). The significant result for *expertise* supports the importance of this criterion, which is among the most mentioned according to [46]. Moreover, significantly higher *expertise* ratings for non-optimized web pages also suggests that not only experts but also laypeople are able to recognize quality differences and identify website operators whose purposes are more educational or informational. Similar to [12], individuals considered objective information such as references and the presence of advertising. Our study showed that the presence of a reference list had a positive effect on multiple criteria, such as trustworthiness and expertise. This result is in line with a study by [53], which found that the presence of references has a positive effect on perceived scientificness and credibility, even when it comes to pseudoscientific texts. One unanticipated finding was that we did not observe differences between the evaluations of laypeople and experts. A possible explanation for this might be that, due to the think aloud task, the subjects took more time to reflect on the respective web pages—an activity that is closely related to the daily work of information professionals. Unexpectedly, the differences for the evaluation criteria were not significant apart from *expertise*. This is surprising, especially for *trustworthiness*, which appears to be quite important in user evaluations [12,46]. One possible explanation is that there is more of a consensus about what constitutes expertise, with evaluations therefore being clearer. *Trustworthiness*, in contrast, can be judged more subjectively, which means that the ratings are rather in a similar spectrum. Furthermore, and contrary to our expectations, this study did not find an influence of subjects' SEO knowledge and attitudes on web page evaluations. This may be explained by the "invisibility" of SEO, since there is no labelling for SEO measures on the SERP, unlike in the case of ads. Even if some users are rather negative towards SEO, it is hardly possible for them to identify SEO measures on a web page and align their evaluations accordingly.

The small sample size of our study posed a limitation. We could only detect significant differences for the criterion *expertise*. A higher number of cases resulting from a greater sample size might have led to clearer results regarding the other criteria. A further limitation concerns the automatic classification of web pages. The rule-based classification, which forms the basis for the classifier algorithm, rests on assumptions by SEO experts. One of the assumptions is that news companies use SEO techniques to increase the visibility of their content [29]. Therefore, it is arguable whether the web page evaluations are solely driven by optimization measures or also by the assumptions within the classifier.

Regarding future research, four starting points present themselves. First, it would be interesting to conduct the study again with a larger sample to obtain more reliable results. This could be achieved by conducting the study more efficiently as a survey-only study without thinking aloud. Second, a more sophisticated automatic classification system could come into play in the repeated study, more precisely indicating the degree of web page optimization. Differences in quality assessment could thus be linked more clearly to the degree of optimization of a specific web page. Third, the transferability of the results to other topics could be investigated; more specifically, whether the findings also apply to other topics relevant to society beyond health information such as political information. Fourth, the study could be repeated with health experts instead of experts in health information assessment, i.e., medical librarians. This would allow a more precise assessment of whether the non-optimized pages are of higher quality than optimized pages.

One implication of our study concerns the issue of visibility of high-quality content on SERPs. Our study has shown that greater expertise is ascribed to non-optimized web pages, which often have no commercial interests. Such websites include those provided by public authorities. In the absence of SEO measures, however, these websites run the risk of being outranked by sites that appear to be of lower quality from the user's point of view but *do* engage in SEO activities. This could lead to users missing out on the higher-quality information. Conversely, one could argue that non-commercially motivated content producers should strengthen their focus on SEO measures so that their content remains or becomes visible to users. This result is in line with the study by Chi et al. [4]. Due to the source selection of laypeople being focused on dot com (rather than dot gov) sites, the authors conclude that websites from governmental and educational institutions "need more promotion." On the other hand, however, the role and responsibility of search engine operators must also be discussed. By changing the ranking algorithm to rely more on trusted sources [47], high-quality offerings from public authorities and other non-commercial providers could benefit in terms of visibility to users.

In our study, we demonstrated that, from the user's point of view, there are quality differences between optimized, often commercial, and non-optimized, often non-commercial, web pages. This highlights the importance of the problem stated in the introduction, namely the potential influence of optimized content on users due to commercial interests of website operators. For future research, we therefore propose to first investigate the extent to which visibility of non-optimized health-related content actually differs from the visibility of optimized content, and second, what concrete impact this has on users' health information seeking behavior.

# 8   Conclusion

In this paper, we presented a study investigating whether quality differences are perceived between health-related web pages that engage in search engine optimization (SEO) and web pages that do not. For this purpose, we conducted a user study consisting of web page evaluations, thinking aloud protocols, and surveys. We acquired *N* = 61 subjects with (experts) or without (laypeople) professional experience in evaluating health information. The automatic classification of the web pages into optimized and non-optimized was carried out using a self-developed SEO classification tool. A manual classification of the source types revealed that optimized pages are much more likely to be commercially motivated (pages of pharmaceutical companies, for instance) than non-optimized pages (pages of public institutions). A higher level of expertise was attributed to non-optimized web pages than to optimized web pages, whereby the latter are ranked higher on the SERPs. The subjects justified their assessments with the more competent and reputable appearance of non-optimized web pages. No differences were found between the web page evaluations of laypeople and experts. This leads us to the following main conclusions. There is a risk that high-quality content, which may not engage in SEO due to a lack of commercial interests, could be outranked by optimized content of lower quality. When making medical decisions, search engine users may thus be missing out on relevant information, potentially leading to uninformed or even harmful decisions. Hence, our study has implications for search engine operators, as they should give priority in their rankings to sources that possess expertise.

## ACKNOWLEDGMENTS

This work is funded by the German Research Foundation (DFG - Deutsche Forschungsgemeinschaft), grant number 417552432. The authors would like to thank Nina Niesche for her help in conducting the survey. All files from the study are available from the OSF repository (https://dx.doi.org/10.17605/OSF.IO/UK4EQ).

# A   APPENDICES
## A.1   Appendix 1

**Table A 1: Survey within the study part "laypeople"**

| No. | Question | Response options and correct answers/coding specifications |
|---|---|---|
| I) Knowledge on SEO | | |
| 1.1 | Do website operators or companies have the ability or influence to appear higher in the Google results list for certain queries without paying any money to Google? | yes (correct), no, I don't know |
| 1.2 | [If "Yes" on question 1.1] Do you know what term is used to describe these measures to improve the ranking in the Google search results list (without payment to Google)? | free input<br>correct: "search engine optimization" or abbreviation/German form<br>partly correct: correct term and at least one incorrect term<br>incorrect: clearly incorrect terms (e.g., ads) |
| 1.3 | [If "Yes" on question 1.1] And by what means can a website be designed or programmed so that it is ranked higher in the Google search results lists? | free input<br>correct: "keywords" or other correct SEO techniques<br>partly correct: correct term and at least one incorrect term<br>incorrect: clearly incorrect SEO techniques (e.g., payment) |
| | Information part "SEO/PSM" | |
| II) Ability to assign results influenceable by SEO/PSM | | |

| No. | Question | Response options and correct answers/coding specifications |
|---|---|---|
| 2.1 | [simple structured SERP, PSM] Are there any search results on this page that can be influenced by the website operator paying Google? | Marking the requested results or skipping the task by specifying that the requested result type is not available on SERP. |
| 2.2 | [simple structured SERP, SEO] Are there any search results on this page that can be influenced by search engine optimization? | Randomized orders of the requested result types (SEO or PSM first) and SERPs (simple or complex first). Example: PSM (simple)->PSM (complex)->SEO (complex)->SEO (simple) |
| 2.3 | [complex structured SERP, PSM] Are there any search results on this page that can be influenced by the website operator paying Google? | |
| 2.4 | [complex structured SERP, SEO] Are there any search results on this page that can be influenced by search engine optimization? | |
| III) Assessments and opinions regarding SEO | | |
| 3.1 | Now please think again about search engine optimization. In your opinion, how strong is the influence of search engine optimization on the ranking of the search results in Google? | Influence of search engine optimization on the order of search results in Google: very strong influence strong influence medium influence little influence no influence I don't know |
| 3.2 | How big are the positive and negative effects of search engine optimization on the Google search results from your perspective? 3.2a) I perceive the positive effects of search engine optimization as ... 3.2b) I perceive the negative effects of search engine optimization as ... | Please mark the appropriate answer in each case: very large large medium low non-existent I don't know |
| IV) Search engine usage | | |
| 4.1 | If you are searching for something online: Which search engine(s) do you usually use? | Bing Ecosia DuckDuckGo Google Web.de Yahoo! Another I don't know/not specified |
| 4.2 | Which search engine do you use most often? | display of all search engines specified in 4.1 |
| 4.3 | Have you ever searched for online health information via search engines? | Yes No I don't know |
| 4.4 | [If "yes" on question 4.3] How often do you search for health information using search engines? | more than once per week about once per week several times per month about once per month less than one per month I don't know |

| No. | Question | Response options and correct answers/coding specifications |
|---|---|---|
| V) Demographic data | | |
| 5.1 | How old are you? | numerical input |
| 5.2 | You are … | female, male, other, prefer not to say |
| 5.3 | Which of the following activities do you mainly pursue? | employee or public official<br>self-employed person, freelancer, entrepreneur<br>student<br>trainee, apprentice<br>pupil<br>housewife/houseman<br>occasionally employed<br>not or no longer employed<br>other |
| 5.4 | What is your highest educational level? | Certificate of Secondary Education without completed apprenticeship<br>Certificate of Secondary Education with completed apprenticeship<br>General Certificate of Secondary Education<br>A-levels<br>University degree<br>None<br>(Still) without school-leaving certificate (e.g., student)<br>Other |